\documentstyle[11pt,newpasp,twoside,epsf]{article}
\def\edcomment#1{\iffalse\marginpar{\raggedright\sl#1\/}\else\relax\fi}
\reversemarginpar

\begin{document}
\title{Direct Simulation of Dense Stellar Systems with GRAPE-6}
 \author{Junichiro Makino}
\affil{Department of Astronomy, University of Tokyo,\\ 7-3-1
Hongo, Bunkyo-ku, Tokyo 113-0033, Japan\\
email:makino@astron.s.u-tokyo.ac.jp \\phone: +81-3-5841-4276, fax: +81-3-5841-7644}

\begin{abstract}
In this paper we describe the current status of the GRAPE-6 project to
develop a special-purpose computer with a peak speed exceeding 100
Tflops for the simulation of astrophysical $N$-body problems. One of
the main targets of the GRAPE-6 project is the simulation of dense
stellar systems. In this paper, therefore, we overview the basic
algorithms we use for the simulation of dense stellar systems and
their characteristics. We then describe how we designed GRAPE hardwares to
meet the requirements of these algorithms. GRAPE-6 will be completed
by the year 2001. As an example of what science can be done on
GRAPE-6, we describe our work on the galactic center with massive
black holes performed on GRAPE-4, the predecessor of GRAPE-6.

\end{abstract}

\setlength{\textfloatsep}{10 pt plus 2 pt minus 2pt}

\section{Introduction}

Direct $N$-body simulation of star clusters or other stellar systems
has proven itself an extremely powerful tool to study the structure
and evolution of stellar systems, since the pioneering work by von
H\"orner (\cite{vonHoerner1960,vonHoerner1963}). The only known way to
do experiments on stellar systems is to construct their models in
computers, because we cannot do laboratory experiments on stellar
systems,

Of course, $N$-body simulation is not the only way to construct
computer models of star clusters. One could use Monte-Carlo (Henon
\cite{Henon1971}, for recent development see Giersz \cite{Giersz1998})
or direct integration of Fokker-Planck (FP) equations (Cohn
\cite{Cohn1980}, Takahashi \cite{Takahashi1996}, Drukier {\it et al.} 
\cite{Drukieretal1999}). In particular, recent advance in the
treatment of the two-dimensional [$f(E,J)$] FP equation made it
possible to use FP code to the study of the evaporation of clusters in
the tidal field of the parent galaxy.

The main advantage of these methods over direct $N$-body simulation is
the calculation cost. While it is still out of reach to perform the
direct simulation of, say, a typical globular cluster with $N=10^6$,
the FP equation is valid in the limit of $N\rightarrow
\infty$. Therefore, when the assumption of large-$N$ limit is good,
methods based on the FP approximation can deliver reliable results for 
the calculation cost orders of magnitudes smaller than that of direct
$N$-body calculation. In principle, one can do the $N$-body simulation
with small $N$ and then try to extrapolate that result to larger
$N$. However, this problem, which is sometimes called ``the scaling
problem'' has proven itself a complex and difficult
problem(Heggie {\it et al.} \cite{Heggieetal1998}). The main reason of the difficulty is
that there are many physical processes of which timescales depend on
the number of particles in different ways, and that there is no way to
adjust all relevant timescales consistently.

On the other hand, methods based on FP approximation do have their own
limitations. For example, it's pretty difficult to extend the
formulation beyond the spherical symmetry. Moreover, they also suffer
the problem of the time scaling, only difference is that they are
affected from the opposite direction. For example, when we want to
include the dynamical effect of the slowly varying potential, FP
approximation goes into trouble. Consider the tidal stripping of
globular clusters with non-circular orbit. The orbital timescale is of
the order of $10^8$ years. The orbital timescale of the stars around
the tidal boundary of the cluster is also $10^8$ years.  On the other
hand, the half-mass relaxation time is of the order of $10^9$
years. Therefore, the timestep to integrate the FP equation is at
the largest $10^8$ years, and in practice much smaller than that. Clearly,
the assumption that the dynamical timescale is smaller than the
Fokker-Plank timestep is broken, and FP calculations tends to grossly
overestimate the escaper rate.

As beautifully demonstrated by Takahashi and Portegies Zwart
(\cite{TakahashiPortegiesZwart1998}), one can incorporate the escaping
timescale into FP calculation. However, in order to do so the model
parameter must be adjusted so that the agreement with reference
$N$-body calculations is achieved. Moreover, this agreement has been
so far achieved only with $N$-body simulations with simple static
``tidal cutoff'', where {\it no} tidal field is imposed and stars
outside the tidal radius are just removed. Whether or not FP
calculations can reproduce the result of $N$-body calculations with
tidal fields is an open question.

The ultimate solution for the scaling problem is to perform $N$-body
simulations with sufficiently large number of particles. In the first
half of this paper, we describe our effort in that direction, the
GRAPE project to develop and use special-purpose computers for
$N$-body simulations. In the last half of this paper, we'd like to
discuss briefly our recent work on $N$-body simulations of galactic
centers with massive black holes, where again the scaling problem
shows up.

\section{Direct simulation of dense stellar systems}

In principle, the direct $N$-body simulation of stellar systems is
very simple and straightforward. The only thing one has to do is to
numerically integrate the following system of equations of motion:
\def\bx{{\bf{ x}}}
\def\bv{{\boldmath v}}
\def\ba{{\boldmath a}}
\def\badot{{\boldmath \dot a}}
\def\batwo{{\boldmath a}^{(2)}}
\def\bathree{{\boldmath a}^{(3)}}
\begin{equation}
 {d^2 \bx_i \over dt^2} = - \sum_{j\ne i}  Gm_j
{\bx_j - \bx_i \over |\bx_j - \bx_i|^3},
\label{eq:basic-equation}
\end{equation}
where $\bx_i$ and $m_i$ are the position and mass of the particle with
index $i$ and $G$ is the gravitational constant. The summation is
taken over all  stars in the system.

In practice, we need  complex  methods and tricks to 
integrate the above equation. Since we cannot cover all important
issues here, we recommend the reviews by Spurzem (\cite{Spurzem1999})
and Aarseth (\cite{Aarseth1999a,Aarseth1999b}). Here, we briefly discuss 
issues directly related to the use of special-purpose hardwares for
$N$-body problem.

As we wrote above, in principle an $N$-body simulation is simple and
straightforward. At each timestep, we calculate the forces on all
particles in the system, and integrate their orbits using some
appropriate integration method. In fact, in Molecular Dynamics
simulation, where one solves the $N$-body problem of atoms interacting
through Coulomb and van der Waals forces, one can rely on this
approach. In astrophysics, however, the nature of the gravitational
interaction makes such approach impractical.

The problem is that the gravitational force is an attractive force
with no characteristic scale length. This fact leads to three
complications. The first one is that the inhomogeneity develops as the
system evolves. In the case of star clusters, this is known as core
collapse or gravitational catastrophe. The central core of the cluster
evolves to higher and higher density, while its mass decreases. The
timestep has to be small enough to integrate accurately the orbits of
particles in the core. Therefore, the timestep decreases as the
cluster evolves. The second one is that even when the core density is
not very high, random close encounters of two particles can lead to
arbitrary short timesteps. The third one is that stable binaries and
more complex hierarchical systems are formed through three-body
encounters and encounters of larger number of particles (such as
binary-binary interactions). These systems have very small orbital
timescales. For a quantitative analysis of these issues, see Makino and
Hut (\cite{MakinoHut1988,MakinoHut1990}).

Fortunately, we can handle these complications by a combination of
techniques. The first two are solved by assigning each particle its
own time and timestep. This scheme, the individual timestep scheme,
was first introduced by Aarseth (\cite{Aarseth1963}), and has been used
for four decades.

In the individual timestep scheme, each particle has its own timestep
$\Delta t_i$ and maintains its own time $t_i$. To integrate the
system, one first selects the particle for which the next time ($t_i +
\Delta t_i$) is the minimum. Then, one predicts its position at this
new time.  Positions of all other particles at this time must be
predicted also. Then the force on that particle from other particles
is calculated. The position and velocity of the particle is then
corrected. The new timestep is also calculated.  The integration
scheme is a variation of Krogh's scheme (Krogh \cite{Krogh1974})
modified for second-order equations. \footnote{Though this scheme is
usually called Krogh's scheme in the field of numerical analysis, to
our knowledge the work by Wielen (\cite{Wielen1967}) is the first to
discuss the implementation of individual timestep with an arbitrary
order integration scheme.}

A modification of this individual timestep algorithm is now used to
achieve higher efficiency on vector machines (McMillan
\cite{McMillan1986}) and special-purpose computers (Makino
\cite{Makino1991a}). This scheme is called the blockstep
scheme. In this scheme, the timesteps of particles are forced to
integer powers of two. In addition, the new timestep of a particle is
chosen so that the time of the particle is an integer multiple of the
new timestep. These two criteria make it possible to force multiple
particles to share exactly the same time. As a result, the efficiency
of vector/parallel hardware is improved significantly.  However, it
should be noted that the average number of particles which share the
same time is not very large, in particular when the central core is
small and dense. Therefore, it is necessary that the force calculation
procedure can achieve reasonable performance when the number of
particles to be integrated in parallel is small.  Since the number of
particles in the core can be as small as 100 or less, it is necessary
that the force calculation procedure can achieve a reasonable speed for
that number.

The third problem, the stable binaries and other hierarchical systems,
is solved essentially by integrating them as small-$N$ subsystem
embedded in the cluster. The motion of the center of mass of the
subsystem and internal motion of its members are separated, and
integrated independently with different timesteps. If the
perturbations from the rest of the system is small, we can ignore
it. In the case of a binary, this means we can completely skip the
numerical integration of the internal motion, since we know the
analytic solution. If the perturbation is not negligible but is small,
we can apply various techniques such as ``Slow KS'' (Mikkola and
Aarseth \cite{MikkolaAarseth1996}).  Here, the hardest problem is not
really the calculation cost. The challenge is how we can recognize
these subsystems and their internal structures, and how we can decide
what method is appropriate for that particular structure.

\section{GRAPE for the simulation of dense stellar systems}

\begin{figure}
\begin{center}
\leavevmode
\epsfxsize = 9cm
\epsffile{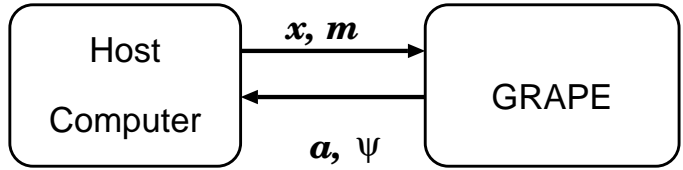}
\caption{Basic idea of GRAPE}
\label{fig:basic-grape}
\end{center}
\end{figure}

\begin{figure}
\begin{center}
\leavevmode
\epsfxsize = 10cm
\epsffile{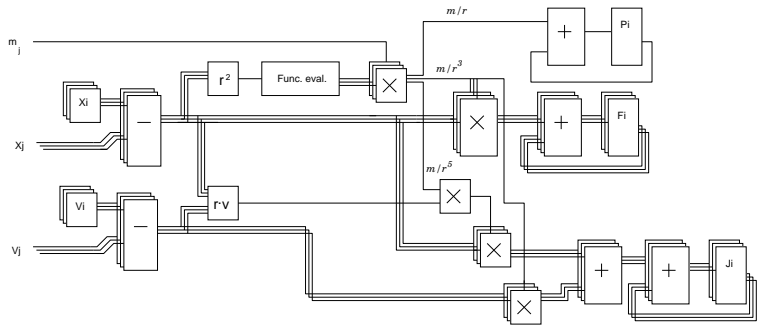}
\caption{Pipelined processor for the calculation of force and its time 
derivative. Reproduced from Makino {\it et al.} 1997 \protect \cite{Makinoetal1997}).}
\label{fig:harppipe}
\end{center}
\end{figure}

We have developed a series of special-purpose computers for $N$-body
simulations, which we call GRAPE (GRAvity PipE). Figure
\ref{fig:basic-grape} gives the basic idea. The host computer, which
is usually a general-purpose workstation running UNIX, send the
positions and masses of particles to the GRAPE hardware. Then GRAPE
hardware calculates the interaction between particles. What GRAPE
hardware calculates is the right hand side of equation
(\ref{eq:basic-equation}). When we use the Hermite scheme
(Makino \cite{Makino1991d2}), the first time derivative of the force must also
be calculated. In this case, velocities of particles are needed.
Figure \ref{fig:harppipe} shows the block diagram of the pipeline unit 
to calculate the force and its time derivative. 

\begin{figure}
\begin{center}
\leavevmode
\epsfxsize = 9cm
\epsffile{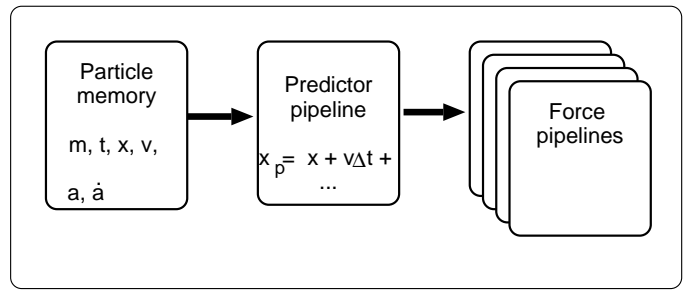}
\caption{GRAPE for individual timestep}
\label{fig:grape-individual}
\end{center}
\end{figure}

In order to combine the individual timestep scheme with GRAPE
hardware, one modification of the basic architecture is
necessary. Figure \ref{fig:grape-individual} shows the change. As
described in the previous section, we have to predict the position
(and velocity  in the case of the Hermite scheme) of all particles
to calculate the forces on the particles in the current
blockstep. This prediction must also be done on the
GRAPE hardware, since otherwize the amount of the calculation the host
computer has to do becomes too large.

In the modified architecture shown in figure
\ref{fig:grape-individual}, the particle memory keeps all data
necessary for prediction, for all particles in the system.  At each
blockstep, the host computer writes the position and velocity of
particles to be updated, and GRAPE calculates the forces on them and
sends the results back to the host. If the number of pipelines is
smaller than the number of particles in the block, this step is
repeated until the forces on all particles in the block are obtained. 
Then the host performs the orbit integration using these calculated
forces, and updates the data of the integrated particles in the
particle memory of the GRAPE hardware.

GRAPE-4 (Makino {\it et al.} \cite{Makinoetal1997}) is the first GRAPE hardware to implement
this modified architecture. In GRAPE-4, one processor board houses one
predictor units and 96 (virtual) force calculation pipelines. The
total system in the maximum configuration consisted of 36 boards
organized into four clusters, and
different boards calculated the forces on the same set of 96
particles. In this way, we met the requirement that the number of
forces calculated in parallel is small, even though the number of
pipelines is large. 

Summation of 9 forces from processor boards in the same
cluster is taken care by the communication hardware, and final
summation of the forces from four clusters is handled by the host.

GRAPE-4 was completed in 1995, and has been used by many researchers
for the study of dense stellar systems. 

\section{GRAPE-6}

In 1997, we started the GRAPE-6 project. It's a five-year project
funded by JSPS (Japan Society for the Promotion of Science), and
the planned total budget is about 500 M JYE.

\begin{figure}
\begin{center}
\leavevmode
\epsfxsize 5.5 cm
\epsffile{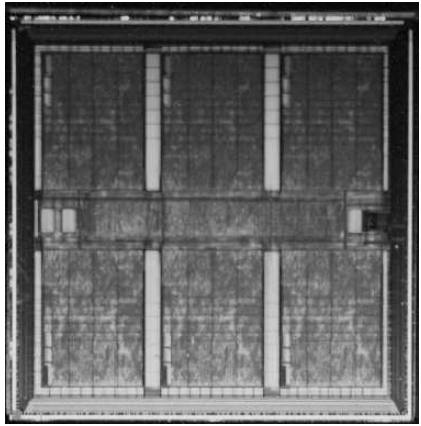}
\caption{The GRAPE-6 processor chip.}
\label{fig:grape6chip} 
\end{center}
\end{figure}

The  GRAPE-6 is essentially a scaled-up
version of GRAPE-4(Makino {\it et al.} \cite{Makinoetal1997}), with the peak speed 
exceeding 100 Tflops. It will consist of around 3000 pipeline
chips, each with the peak speed of 40 Gflops. In comparison, GRAPE-4
consists of 1700 pipeline chips, each with 600 Mflops. The increase of
a factor of 100 in speed is achieved by integrating six pipelines into
one chip (GRAPE-4 chip has one pipeline which needs three cycles to
calculate the force from one particle) and using 3--4 times higher
clock frequency. The advance of the device technology (from $1\mu {\rm
m}$ to $0.25\mu {\rm m}$) made these improvements possible. Figure
\ref{fig:grape6chip} shows the first sample of the processor chip
delivered in early 1999. The six pipeline units are visible.

\begin{figure}
\begin{center}
\leavevmode
\epsfxsize 9.5 cm
\epsffile{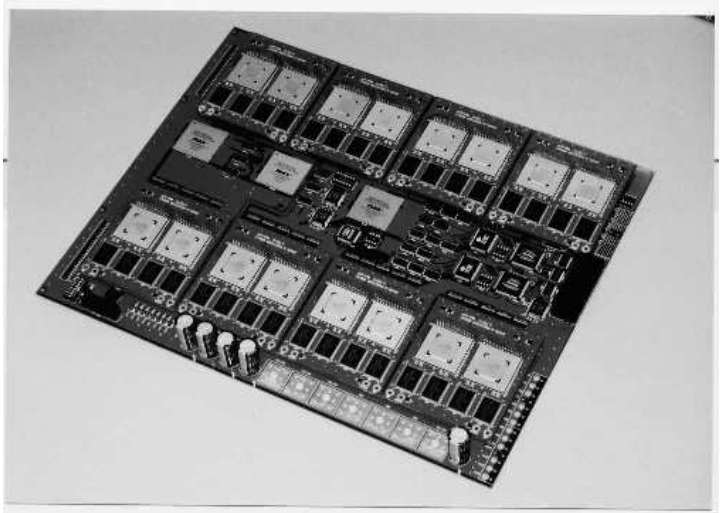}
\caption{The processor board of the GRAPE-6 with 16 processor
chips. Two processor chips are mounted on modules, on which four
memory chips are also mounted. One board houses eight modules.}
\label{fig:grape6board} 
\end{center}
\end{figure}

\begin{figure}
\begin{center}
\leavevmode
\epsfysize 9 cm
\epsffile{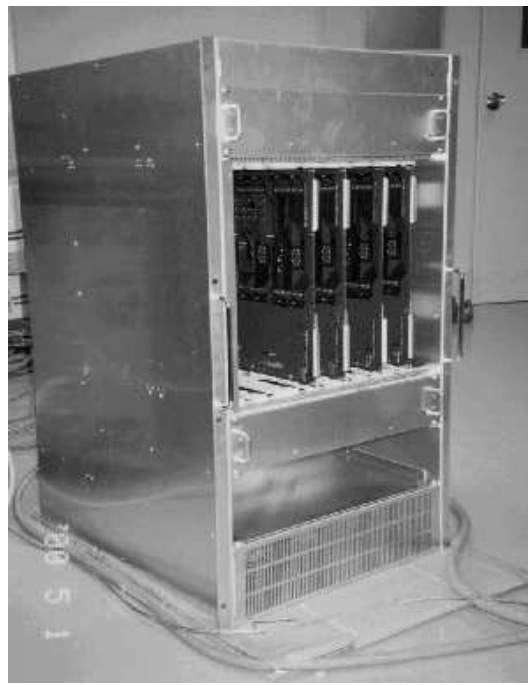}
\caption{The prototype system with four processor board.}
\label{fig:grape6box} 
\end{center}
\end{figure}

Figures \ref{fig:grape6board} and \ref{fig:grape6box} shows the
processor board with 16 processor chips and the prototype four-board
system. This four-board system has the theoretical peak speed of 2.1
Tflops, and has achieved the sustained speed of 1.1 Tflops for the
simulation of 1 million-body system.

GRAPE-6 will be completed by the year 2001. We plan to make small
version of GRAPE-6 (peak speed of around one teraflops) commercially
available by that time. We've found that the commercial availability
of small machines is essential to maximize the scientific outcome from
GRAPE hardwares.

Compared to GRAPE-4, GRAPE-6 will give us 100 times more computer
power. For simulation of star clusters for the relaxation timescale,
this means a factor of five increase in the number of particles we can 
handle. For short simulations, the increase would be a factor of
10. In the case where we can use tree algorithms, in principle a
factor of 100 increase is possible if the host computer has
a sufficiently large memory. Table 1 gives a rough idea of what is
currently possible with GRAPE-4 and what will be possible soon with
GRAPE-6.

\begin{table}
\begin{center}
\caption{Particle Number of Simulation Feasible on GRAPE-4 and 
6}

\smallskip

\begin{tabular}{lcc}
\tableline
Problem Aria &  GRAPE-4 &  GRAPE-6 \\
\tableline
Planet Formation  & $5\times10^4$  & $10^6$ \\
Star  Cluster  & 5$\times10^4$  & 3$\times10^5$ \\

Black Hole Binary in Galactic Nucleus& 10$^6$  & $10^7$ \\

Galaxy Evolution \&  Interactions & 2$\times10^6$  & 10$^8$\\
Large Scale Structures  & 3$\times10^7$ & 3$\times10^9$ \\
\tableline
\tableline
\end{tabular}
\end{center}
\end{table}

\section{BH binaries in galactic cores}

In this section, we briefly describe our resent work on the massive
central black holes in the centers of galaxies (Makino and Ebisuzaki
\cite{MakinoEbisuzaki1996}, Makino \cite{Makino1997}, Nakano and
Makino \cite{NakanoMakino1999a,NakanoMakino1999b}).  The problem is
what happens when two galaxies, each with a central black hole, merge. 
The merging of two galaxies is a rather common event. According to the
standard scenario of the structure formation in the universe
(hierarchical gravitational clustering), galaxies are constructed
``bottom up'' from smaller objects through merging. On the other hand,
almost all galaxies except for some dwarfs seem to have central black
holes. Thus, if two such galaxies merge, the merger remnant would have
two black holes. These black holes would settle in the center of the
merger remnant, and would eventually form a binary. Two questions
arise: (a) What is the structure of the central region of the merger
galaxy with two black holes? and (b) What will happen to the binary
itself? Will two black holes eventually merge?

Makino and Ebisuzaki (\cite{MakinoEbisuzaki1996}) studied the first
problem, assuming that black holes would eventually merge. They
performed the simulation of hierarchical (repeated) merging similar
to those by Farouki {\it et al.} (\cite{Farouki1983}), but with central
black holes. What Farouki {\it et al.} found is that the core radius
remains almost unchanged after merging, even though the half-mass radius
almost doubles after each merging event. This result is consistent
with the theoretical prediction based on the conservation of the
central phase space density, but in complete contradiction with the
observations of luminous elliptical galaxies. Ground-based
observations in 1980s demonstrated that there is strong correlation
between the effective radius and the ``core'' radius of luminous
ellipticals (Lauer \cite{Lauer1985}). HST observations (Gebhaldt {\it
et al.} \cite{Gebhardtetal1996})
have shown that the ``cores'' are actually all cusps with the profile
$\rho = r^{-0.2 \sim -1}$, where $\rho$ is the volume luminosity
density.

\begin{figure}
\begin{center}
\leavevmode
\epsfxsize = 6.3cm
\epsffile{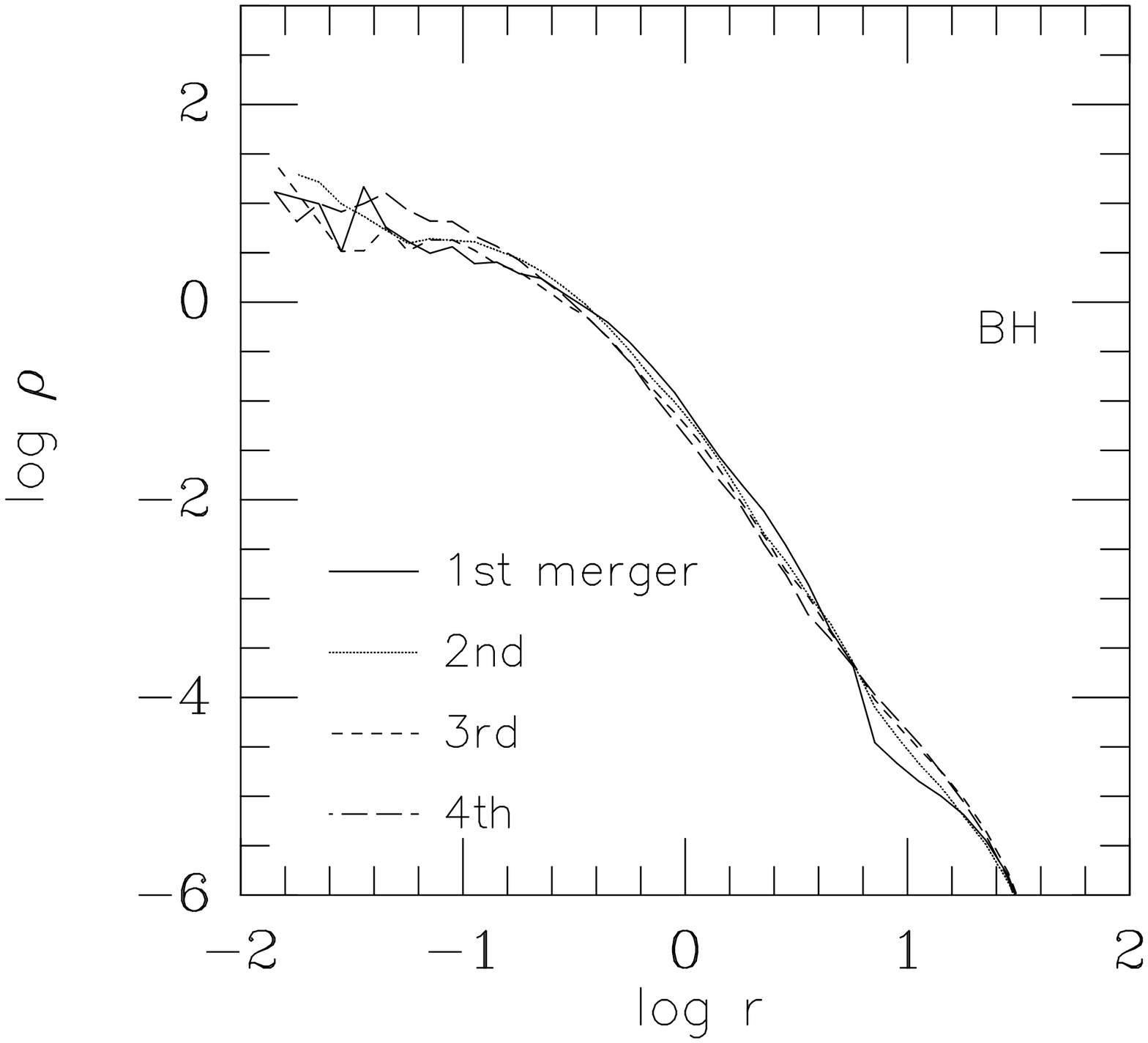}
\epsfxsize = 6.3cm
\epsffile{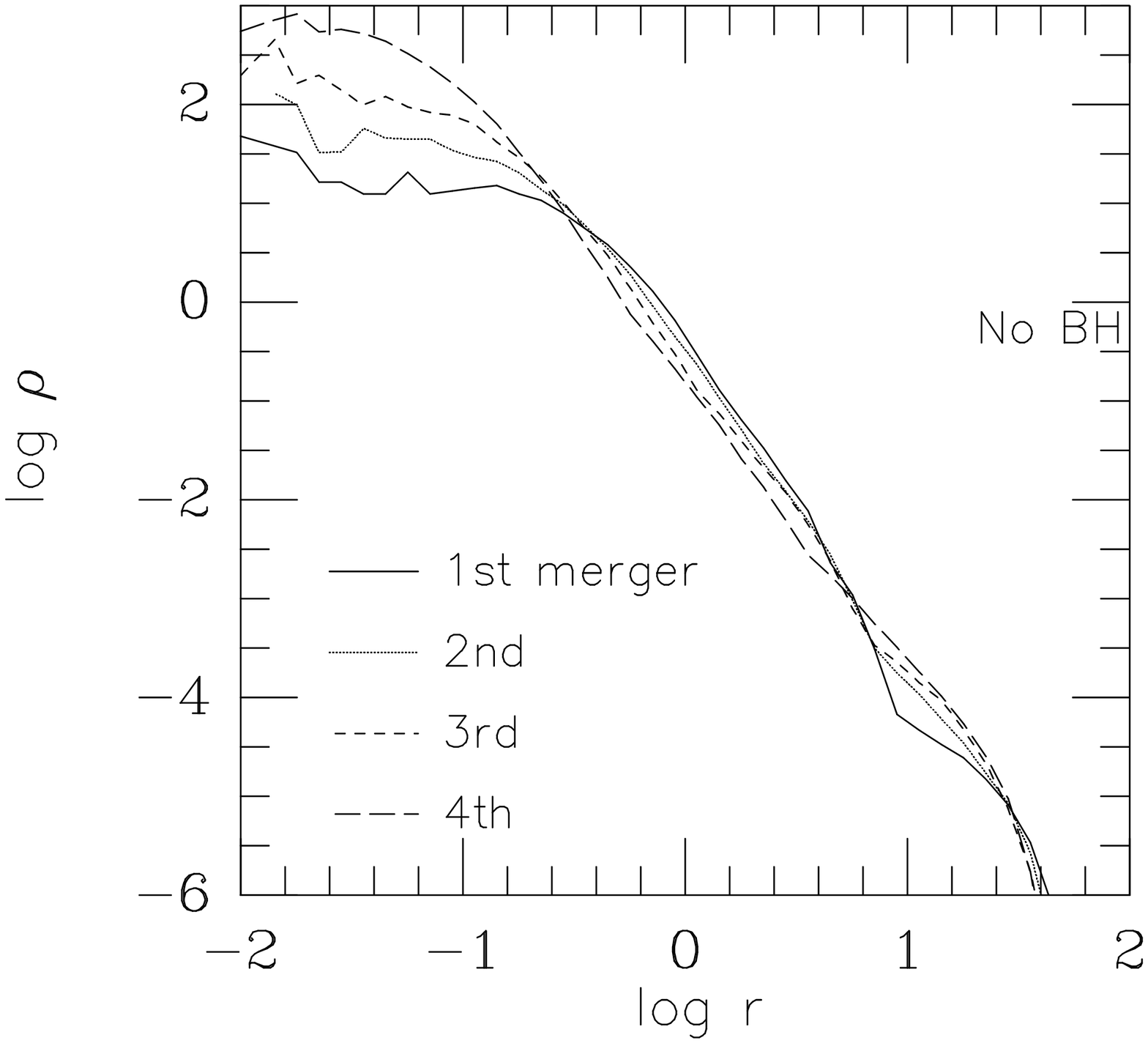}
\caption{The density profiles of the mergers with central black
holes. Profiles are scaled so that the half-mass radii are the same
for all remnants. The panel in the left side shows the results of runs 
with central black holes, and that in the right without black holes. Reproduced from Makino and Ebisuzaki (1997).}
\label{fig:makinoebisu}
\end{center}
\end{figure}

Figure \ref{fig:makinoebisu} show the result of simulations with
GRAPE-4. The central region of the merger with central black hole
looks like the shallow cusps observers found in luminous ellipticals,
and the radius of this cusp region is a constant fraction of the
half-mass radius. Thus, numerical simulations of merging of galaxies
with black holes reproduced the observed characteristic of the central 
regions of luminous ellipticals rather well. 

Nakano and Makino (\cite{NakanoMakino1999a,NakanoMakino1999b})
investigated why the cusp is formed. First, they tried to understand
the formation mechanism better by means of simplified experiments, in
which just one black hole was let to sink to the center of a
galaxy. Except for the case that the black hole is originally in the
center, the cusp with $\rho = r^{-0.5}$ is universally formed, and the 
mass within the cusp region is comparable to that of the mass of the
central BH.

In the second paper (Nakano and Makino 1999b), they came
up with a simple explanation for the formation mechanism. The
black hole sink to the center through the dynamical friction, in the
dynamical timescale. When the black hole settled at the center of the
galaxies, almost no star was strongly bound to the black hole. This is
because the black hole sinks to the center relatively slowly. The
sinking timescale is not so slow that the adiabatic approximation can
be applied to the binding energy of the stars, but sufficiently slow that 
the binding energy of most of the stars does not change very much.
Thus, the distribution function $f(E)$ has a sharp cutoff at the
energy close to the central  potential depth of the initial galaxy
model.

If there is a central massive object and $f(E)$ has a sharp cutoff, we
can show that the central region is a cusp with the slope $-0.5$ as
follows. 

The density is obtained by integrating the distribution function in
the velocity space as follows:
\begin{equation}
\rho(r)=4\pi \int^{0}_{\psi(r)} f(E) \sqrt{2|E-\psi(r)|} d{E},
\label{eq:density}
\end{equation}
where $\psi(r)$ is the potential at distance $r$ from the center,
where the black hole lies. Note that we assumed that the velocity
distribution is isotropic.

We assume that $f(E) = 0$ if $E<-E_0$. For any $r$ with $\psi(r) <
E_{\rm 0}$, equation (\ref{eq:density})
can be rewritten as
\begin{equation}
\rho(r)=4\pi \int^{0}_{E_0} f(E) \sqrt{2|E-\psi(r)|} d E,
\end{equation}
which can be expanded as
\begin{equation}
\rho(r) 
 = 4\sqrt{2}\pi{\sqrt{|\psi(r)|}} \int^{0}_{E_{0}} f({E}) \left[%
1-\frac{1}{2}\frac
 {{ E}}{\psi(r)}+O\left({\left[\frac{{E}}{\psi(r)}\right]}^{2} \right)
 \right] d{ E},
\end{equation}
Therefore, we have
\begin{equation}
\rho(r)  \propto  \sqrt{|\psi(r)|} \sim \sqrt{\frac{G M_{\rm BH}}{r}}.
\end{equation}

For the fate of the black hole binary, we performed simulations with
very wide range of the number of particles (2,048 to 262,144) to investigate 
the dependence of the growth timescale of the binding energy of the
binary on the number of particles in the galaxy. What we found was 
intriguing. The timescale seemed to be proportional to $N^{1/3}$,
while, theoretically, the timescale must be proportional to $N$, since 
the timescale should be limited by the timescale of filling the loss
cone through two-body relaxation. We have not yet understood why our
numerical experiments did not agree with the theoretical
prediction. Simulations with larger $N$, which will be possible with
GRAPE-6, will give us important clues for the understanding of this
problem.

\section{Discussion}

In this paper, we briefly overviewed the computational issues of
direct $N$-body simulations and the use of special-purpose computers,
and we reviewed the plan and development status of GRAPE-6. We believe
GRAPE-6 will give us many new important scientific results, much in
the same way as GRAPE-4 have done so in the five years since its
completion.

Since GRAPE-6 is now close to its completion, it might be worthwhile
to speculate on what will come next. The silicon device technology has
been advancing as usual, and will do so at least for one more decade
or two. If we will start the development of the new system just after
the completion of GRAPE-6, we will use the technology safely available
by 2002, which is either 0.13 or $0.09 \mu{\rm m}$ technology. This
will give us around a factor of six increase in the transistor
count. For the clock speed, we expect that the improvement by a factor of
three will not be too hard, mainly because the current clock of GRAPE-6
chip is not very high. Thus, we can achieve about a
factor of 20 improvement in the speed over that of GRAPE-6 chip, reaching around 0.7 Tflops per
chip. With three thousand chips, the peak speed would reach two
Petaflops.

The target number of particles for such system would be around 1
million, for star cluster simulations. For that number, however, one
might ask whether or not the direct force calculation is really a
good approach, even though it guarantees very good cost-performance
ratio.

There are two  widely used algorithms to reduce the
calculation cost. One is the neighbor scheme (Ahmad and Cohen
\cite{AhmadCohen1973}), and the other is the treecode (Barnes and Hut
\cite{BarnesHut1986}). For general-purpose computers, both algorithms
give pretty large improvement in the overall calculation speed. The
theoretical gain of the neighbor scheme is $O(N^{1/4})$ (Makino and
Hut \cite{MakinoHut1988}), and the break-even point is as small as
$N=25$. For the treecode, the gain is $O(N/\log N)$. The break-even
point depends on the required accuracy, but $N=10^6$ is certainly
large enough to guarantee a pretty large gain.

However, we (or anybody else) have not yet conceived a feasible way to
combine the neighbor scheme and GRAPE, or treecode with individual
timestep and GRAPE. In both cases, the problem is that forces on
different particles are calculated differently. As a result, $n$
pipelines which calculate the forces on $n$ different particles need
$n$ different data from the particle memory, while in the case of the
simple direct summation all pipelines can use the same data. In
principle, if each pipeline has its own memory, we can reconcile
massively parallel GRAPE and these algorithms. With the
next-generation system, it will be necessary to integrate the memory
and the pipeline processor anyway, since otherwize we cannot achieve
required memory bandwidth even for the case of the simple GRAPE
architecture. This integrated memory and pipeline might give us the
opportunity to implement more efficient force calculation algorithms.

\section*{Acknowledgments}

We would like to thank Daiichiro Sugimoto, Toshikazu Ebisuzaki, Makoto
Taiji, Tomoyoshi Ito, Toshiyuki Fukushige and many others who were involved in the
development of the six generations of GRAPE hardwares, and Yoko
Funato, Simon Portegies Zwart, Steve McMillan, Piet Hut and again many
others for discussions and collaborations in software development and
scientific works on GRAPE hardwares.  This work is supported by the
Research for the Future Program of Japan Society for the Promotion of
Science (JSPS-RFTP97P01102).

\def\pasj{PASJ}
\def\mn{MNRAS}
\def\MN{MNRAS}
\def\mnras{MNRAS}
\def\apj{ApJ}

\newcommand{\etalchar}[1]{$^{#1}$}
\newcommand{\noopsort}[1]{} \newcommand{\printfirst}[2]{#1}
  \newcommand{\singleletter}[1]{#1} \newcommand{\switchargs}[2]{#2#1}

\end{document}